\documentclass[11pt]{article}

\usepackage{amsmath,amssymb}
\usepackage{slashed}
\usepackage{xcolor} 
\usepackage{graphicx}
\usepackage{dcolumn} 
\usepackage{bm} 
\usepackage{epsfig}
\usepackage{epstopdf}
\usepackage{grffile}
\usepackage{color}
\usepackage{colordvi}
\usepackage{amsmath,amssymb}
\usepackage{rotating}
\usepackage{lscape}
\usepackage{cite}
\usepackage{float}
\usepackage{hyperref}
\usepackage{url}
\usepackage{graphicx}
\usepackage{color}
\usepackage{amssymb}
\usepackage{amsmath}
\usepackage{mathtools}
\usepackage{bm}	
\usepackage{flexisym}
\usepackage{amssymb}
\usepackage{physics}
\usepackage{mathrsfs}
\usepackage{simplewick}
\usepackage{tikz}
\usetikzlibrary{arrows,shapes}
\usetikzlibrary{trees}
\usetikzlibrary{matrix,arrows} 			
\usetikzlibrary{positioning}			
\usetikzlibrary{calc,through}			
\usetikzlibrary{decorations.pathreplacing}  
\usepackage{pgffor}						
\usetikzlibrary{decorations.pathmorphing}
\usetikzlibrary{decorations.markings}
\tikzset{
	vector/.style={decorate, decoration={snake}, draw},
	provector/.style={decorate, decoration={snake,amplitude=2.5pt}, draw},
	antivector/.style={decorate, decoration={snake,amplitude=-2.5pt}, draw},
	fermion/.style={draw=black, postaction={decorate},
		decoration={markings,mark=at position .55 with {\arrow[draw=black]{>}}}},
	fermionbar/.style={draw=black, postaction={decorate},
		decoration={markings,mark=at position .55 with {\arrow[draw=black]{<}}}},
	fermionnoarrow/.style={draw=black},
	gluon/.style={decorate, draw=black,
		decoration={coil,amplitude=4pt, segment length=5pt}},
	scalar/.style={dashed,draw=black, postaction={decorate},
		decoration={markings,mark=at position .55 with {\arrow[draw=black]{>}}}},
	scalarbar/.style={dashed,draw=black, postaction={decorate},
		decoration={markings,mark=at position .55 with {\arrow[draw=black]{<}}}},
	scalarnoarrow/.style={dashed,draw=black},
	electron/.style={draw=black, postaction={decorate},
		decoration={markings,mark=at position .55 with {\arrow[draw=black]{>}}}},
	bigvector/.style={decorate, decoration={snake,amplitude=4pt}, draw},
}
\tikzstyle{block} = [draw, rectangle, minimum height=3em, minimum width=6em]
\usepackage{hyperref}

\usepackage{titling}
\usepackage{subcaption}
\newcommand{\subtitle}[1]{%
	\posttitle{%
		\par\end{center}
	\begin{center}\large#1\end{center}
	\vskip0.5em}%
}
\begin{document}

\begin{center}

\vspace*{15mm}
\vspace{1cm}
{\Large \bf Prospects for probing axionlike particles at a future hadron collider through top quark production}

\vspace{1cm}

{\bf Yasaman Hosseini  and Mojtaba Mohammadi Najafabadi }

 \vspace*{0.5cm}

{\small\sl 
School of Particles and Accelerators, Institute for Research in Fundamental Sciences (IPM) P.O. Box 19395-5531, Tehran, Iran } \\

\vspace*{.2cm}
\end{center}

\vspace*{10mm}

%
%
\begin{abstract}\label{abstract}
Axionlike particles (ALPs) emerge from
spontaneously broken global symmetries 
in high energy extensions of the Standard Model (SM).
This causes ALPs to be among the objectives of future experiments 
which intend to search for new physics beyond the SM.
We discuss the reach of future pp collider FCC-hh
in probing the ALP model parameters through top quark pair production associated 
with ALP ($t\bar{t}+\text{ALP}$) in a model-independent approach. 
The search is performed in the semileptonic decay mode of $t\bar{t}$ and
the analysis is performed using a parametric simulation
of the detector response for a projected integrated luminosity of $\rm 30~ab^{-1}$.
It is shown that $t\bar{t}+\text{ALP}$ production at the FCC-hh is a promising channel with significant
sensitivity to probe the ALP coupling with gluons. The ALP coupling with gluons obtained
 from HL-LHC and other experiments are presented for comparison.
\end{abstract}

\newpage

\section{Introduction}

Axion-like particles (ALPs) are pseudo-Goldstone bosons, that can appear from the spontaneous breaking
of some global symmetries at energy scales well above the electroweak scale. 
In recent years, there has been much interest in ALPs because of their various worthy
aspects. ALPs possess many applications based on their masses and couplings in
the parameter space. 
 ALPs can solve the strong CP problem \cite{r1} and they are proper candidates for non-thermal Dark Matter (DM) \cite{r2}. 
ALPs can play a vital role in baryogenesis giving an explanation for the observed imbalance in matter and antimatter \cite{r3}
and are able to explain the neutrino mass problem through an ALP-neutrino interaction which causes neutrinos 
to earn mass \cite{r4}. Furthermore, ALPs can address the muon anomalous magnetic dipole moment \cite{r5}
and the excess  observed in the rare $K$ mesons studies reported by the
KOTO experiment \cite{r6}. 

ALPs are mostly probed in a model-independent effective field theory (EFT) framework. 
The strength of ALPs couplings  to SM fields is proportional to the inverse of 
$U(1)$ spontaneous symmetry breaking scale $f_{a}$, which is much higher than the
 electroweak symmetry breaking scale of the SM.
So far, a remarkable region of the ALP parameter space
in terms of its mass and couplings has been probed or will be studied by cosmological observations, low-energy experiments, 
and collider searches \cite{Fukuda:2015ana, Izaguirre:2016dfi, Dobrich:2015jyk, Carmona:2022jid,l1,l2, BREAD:2021tpx, ATLAS:2020hii, CMS:2018erd, Preskill:1982cy, Abbott:1982af, Dine:1982ah}.

Very light ALPs with masses below the electron pair mass ($m_{a} < 2m_{e}$)
are only allowed to decay into a pair of photons. Based on the ALPs masses and couplings, heavier ones
are allowed to decay into hadrons and charged leptons. 
The decay rates of light ALPs are usually very small such that they can travel a long distance before they decay.
Long-lived ALPs show up as invisible particles at colliders therefore they appear as missing energy in the detectors
since they decay outside the detector environment.  There are several proposals for searches at collider experiments 
to probe long-lived ALPs via mono-jet, mono-$V$ ($V = \gamma,W,Z$), and jet$+\gamma$ \cite{x1,x2,x3,x4,x5,mojtaba, haghighat, Bauer:2018uxu}.
Searches for ALPs via exotic Higgs decays $H\rightarrow Z+a$ and $H\rightarrow a + a$ with ALP decays to diphoton and dilepton
at the LHC have provided remarkable sensitivities in a vast region of parameter space \cite{l1,l2,l3, ATLAS:2021kxv, ATLAS:2021hbr}. 
 There are searches for ALPs through the production of dijet in association with an ALP and jet+ALP
at the LHC and FCC-hh which can be found for instance in Refs. \cite{x1,haghighat}.
It has been shown that dijet+ALP channel using multivariate analysis  
provides a strong sensitivity to the ALP coupling with gluons.  
Although the bounds on the ALP coupling with gluons from dijet+ALP and jet+ALP  \cite{haghighat} 
are very strong, it is worth performing complementary searches through $t\bar{t}+ALP$. 
Furthermore, the structure of the fermionic ALP couplings is specific as it consists of the Yukawa matrices,
 as a result, the ALP is expected to couple more strongly to third generation quarks. This makes $t\bar{t}+ALP$ an important channel to explore the ALP model.
 
In this paper, we propose a search for strong and fermionic couplings of ALP through the associated production of an ALP 
with a pair of $t\bar{t}$ in proton-proton collisions at the future circular collider (FCC-hh) \cite{fcchh} at 
a center-of-mass energy of $100$ TeV. In particular, the concentration is on a region of the parameter space,
in which ALP does not decay inside the detector and manifest as missing energy.

This paper is organized as follows. In section \ref{framework}, 
an introduction to the ALP model is presented.
Section \ref{ttalp}  is dedicated to present the details of search 
for the ALP model using $t\bar{t}+ALP$.  In section \ref{sum}, a summary
 of the results and discussions are given.

\section{Effective Lagrangian for axionlike particles}
\label{framework}

The theoretical framework adopted throughout this work
is a linear effective field theory  where
electroweak physics beyond the SM is expressed
by a linear EFT expansion versus gauge invariant operators ordered by their mass
dimension. The model includes  SM plus an
ALP where the scale of the new physics is the
ALP decay constant $f_{a}$. 
The most general effective Lagrangian describing ALP interactions with SM fields 
up to dimension $D=5$ operators has the following form~\cite{x1}:
\begin{eqnarray}
\begin{split}
	\mathcal{L}_{eff}^{D\leq5} 
	&= 
	\mathcal{L}_{SM} + 
	\frac{1}{2} (\partial^\mu a)(\partial_\mu a)
	- \frac{1}{2} m_a^2 a^2 + c_{a \Phi} \mathbf{O}^\psi_{a\Phi}
	 \\
	&-  c_{gg} \frac{a}{f_a} G^{A}_{\mu\nu} \tilde{G}^{\mu\nu,A} 
	-  c_{WW} \frac{a}{f_a} W^{A}_{\mu\nu} \tilde{W}^{\mu\nu,A}
	-  c_{BB} \frac{a}{f_a} B_{\mu\nu} \tilde{B}^{\mu\nu},
\end{split}
\label{bosonic11_Lagrangian}
\end{eqnarray}
where
\begin{eqnarray}
\mathbf{O}^\psi_{a\Phi}
\equiv 
i\left(\bar{Q}_L \mathbf{Y}_U\tilde\Phi u_R-\bar{Q}_L \mathbf{Y}_D\Phi d_R-\bar{L}_L\mathbf{Y}_E\Phi e_R\right)\frac{a}{f_a} + \text{h.c.}
\label{ALP-Yukawa}
\end{eqnarray}
where  $e_R$,$d_R$,$u_R$ are $SU(2)_L$ singlets and $L_{L}$ and $Q_{L}$ are the $SU(2)_{L}$ doublets.  
The ALP EFT Lagrangian of Eq.\ref{bosonic11_Lagrangian} is implemented in  \texttt{FeynRules} \cite{Alloul:2013bka} 
according to the notation of Ref.\cite{x1}.  The obtained Universal FeynRules Output (UFO) model \cite{Degrande:2011ua} \footnote{\url{http://feynrules.irmp.ucl.ac.be/attachment/wiki/ALPsEFT/ALP_linear_UFO.tar.gz}} 
 is embedded to \texttt{MadGraph5\_aMC@NLO}~\cite{Alwall:2011uj} 
to compute the cross sections and to generate the ALP signal events.

\subsection{ALP decays}

According to the ALP  interactions presented by the effective Lagrangian Eq. \ref{bosonic11_Lagrangian},  ALP
is allowed to decay into pairs of SM particles. 
For the MeV-scale ALPs, the decays into photons, charged leptons and
light hadrons are dominant.
The diphoton decay mode is the most important one for light ALPs with mass $m_{a} < 2m_{e} = 1.022$ MeV.
As $m_{a}$ increases to $2m_{e}$ and above, the leptonic decay modes $a \rightarrow l^{+}l^{-}$ becomes accessible. 
The ALP hadronic decay modes appear when $m_{a} > m_{\pi}$ and arise from 
the ALP decays $a\rightarrow gg$ and $a\rightarrow q\bar{q}$.
The triple pion decay modes $a\rightarrow \pi^{+}\pi^{-}\pi^{0}$ and $a\rightarrow \pi^{0}\pi^{0}\pi^{0}$ 
are the main hadronic modes for $m_{a} < 1$ GeV. Other ALP hadronic decay modes such as $a\rightarrow \pi^{0}\gamma\gamma$ and
$a\rightarrow \pi^{+}\pi^{-}\gamma$
are suppressed with respect to $\pi^{+}\pi^{-}\pi^{0}$ and $3\pi^{0}$ due to the presence of powers of the fine structure constant \cite{x3}.

One should note that a fraction of ALPs decay inside the detector environment and consequently they do not appear 
as missing energy.  The decay length of ALP $\mathcal{L}_{a}$  is proportional to $ \sqrt{\gamma^{2}-1}/\Gamma_{a}$, where 
$\Gamma_{a}$ and $\gamma$ are the total width and ALP Lorentz factor, respectively. 
The ALP decay probability in the detector volume is proportional to $e^{-\mathcal{L}_{\text{det}}/\mathcal{L}_{a}}$.
$\mathcal{L}_{\text{det}}$ is the transverse distance of the detector component from the collision
point. In this study, the probability that the ALP does not decay inside the detector 
and escapes detection is considered event-by-event. The total width of ALP is obtained from Ref.\cite{yotam}
where the chiral perturbation theory and vector meson dominance model have been used in width calculations.
For instance, the decay probability for an ALP with $m_a = 10$ MeV and $|\vec{p}_a|$ = 242 GeV
is 0.0053 while the decay probability for an ALP with $m_a = 70$ MeV and $|\vec{p}_a|$ = 387 GeV is 0.999.

\section{ALP production associated with a pair of top quark}
\label{ttalp}

\begin{figure}[h]
\centering
\includegraphics[width=5.5 cm]{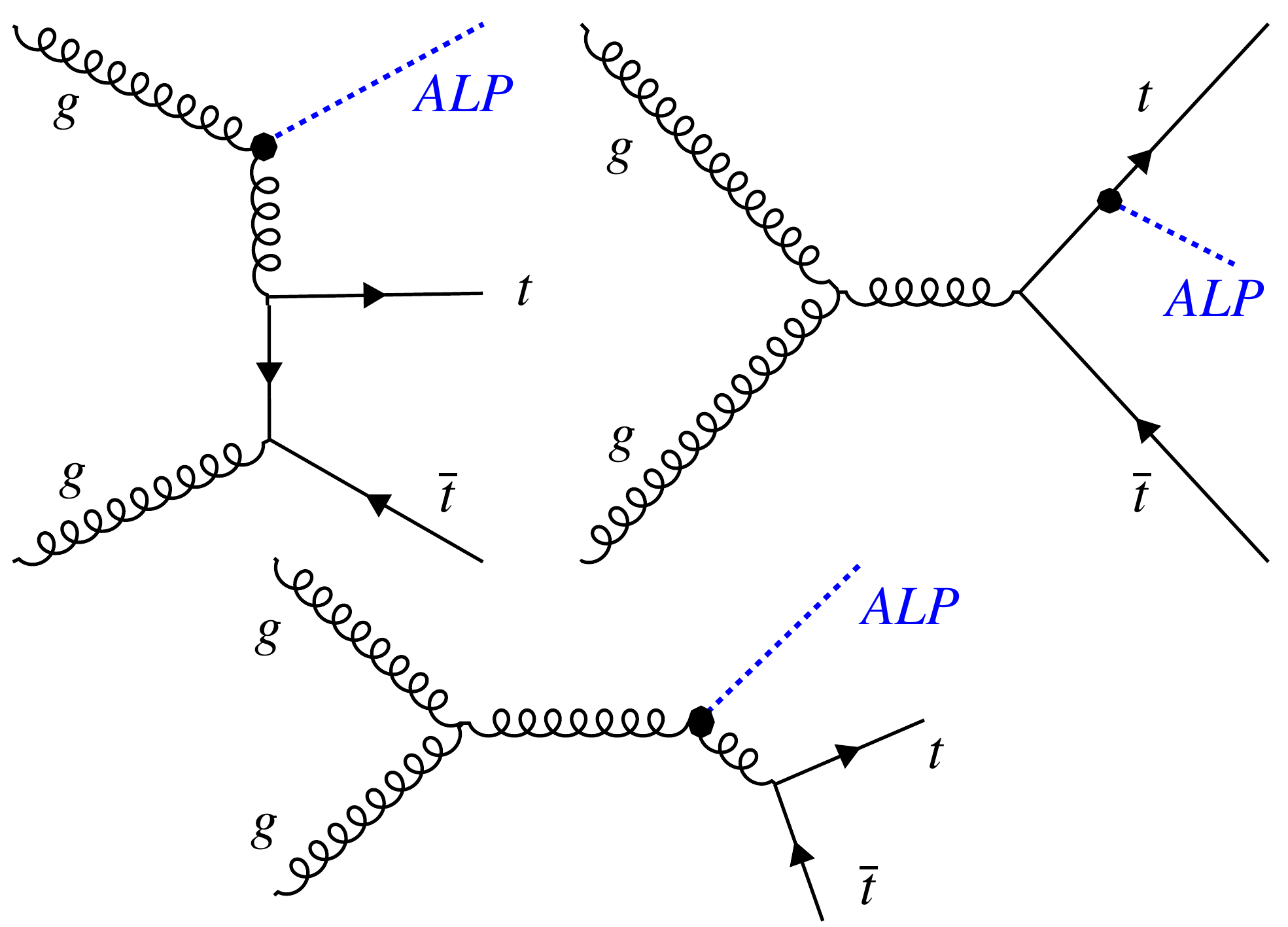}
\caption{Representative leading order Feynman diagrams for 
production of a pair of top quark with an ALP in pp collisions.\label{fig1}}
\end{figure}

Top quark pair production in association
with an ALP in proton-proton collisions at a center-of-mass energy 
of 100 TeV is used to probe the parameter space of the ALP model.
As indicated previously, the focus is on the ALPs that do not decay  within the detector volume and are not detected by
the detectors appearing as missing momentum. 
Figure \ref{fig1} depicts the representative Feynman diagrams
for $t\bar{t}+ALP$ in proton-proton collisions.

At production level, this process is  sensitive to
 $c_{gg}$ and $c_{a\Phi}$. Assuming one non-vanishing ALP coupling 
at a time, the cross sections $\sigma(pp \rightarrow t\bar{t}+ALP)(c_{XX})$ at leading order (LO)
read:
\begin{eqnarray}
&&    \sigma({c_{gg}}) = 459.6  \Bigl(\frac{c_{gg}}{f_a}\Bigr)^2 \text{pb},   \nonumber \\  
&&    \sigma({c_{a\Phi}}) = 2.45  \Bigl(\frac{c_{a\Phi}}{f_a }\Bigr)^2  \text{pb},
    \label{cs}
\end{eqnarray}
where $f_{a}$ is in unit of TeV and the cross sections are calculated using \texttt{MadGraph5\_aMC@NLO}
with the NNPDF23 \cite{Ball:2012cx} as the parton distribution functions (PDFs) of proton.  
The cross sections are obtained for the value of ALP mass $m_{a} = 1$ MeV and 
change up to $10\%$ when $m_{a}$ increases to 100 MeV.
This is expected as in this mass range,
$m_{a}$ is negligible in comparison to the typical energy scale of the process.
The total cross section of the SM $t\bar{t}$ production at leading order calculated by \texttt{MadGraph5\_aMC@NLO}
is 24673.5 pb.
From Eq.\ref{cs}, it is clear that there is more sensitivity to $c_{gg}$ than $c_{a\Phi}$
which is due to the fact that $c_{gg}$ appears in both initial and final states  
and the large gluon PDF. Since the $t\bar{t}+ALP$ rate has no significant sensitivity on $c_{a\Phi}$ coupling
with respect to $c_{gg}$, weaker bound on $c_{a\Phi}$ is expected.

As ALP escapes detection, the $t\bar{t}+ALP$ can be probed through the $t\bar{t} + E_{T}^{miss}$ signature.
Similar signature has been studied by the CMS and ATLAS collaborations in Refs.\cite{CMS:2017dcx, ATLAS:2017hoo} 
to explore simplified models for dark matter where a mediator exists
that couples to both the SM particles and dark matter. 
These studies investigate the production of a fermionic dark matter  through
 a color-neutral scalar or pseudo-scalar particle ($\phi$) exchange
 where the couplings between the new (pseudo)scalar and SM particles are Yukawa like.
Therefore, the mediator is expected to be produced mainly in association with heavy quarks
or through loop induced gluon-gluon fusion.
The distinctive signature for dark matter in $t\bar{t}+\phi$ production followed by $\phi \rightarrow \chi\chi$, where $\chi$ is dark matter field,
is the emergence of a high missing transverse momentum recoiling against $t\bar{t}$ system.

In this analysis, the same as the past LHC search \cite{mojtaba, CMS:2017dcx},
the concentration is on the semi-leptonic $t\bar{t}$ decay channel and follow similar selection.
The final state consists of one charged lepton, four jets, and large missing transverse momentum.
The main background sources to signal arise from $t\bar{t}$, $W+jets$, $Z+jets$, single top production,
and diboson. 
 All background contributions are estimated from simulation.
Both signal and background processes are generated using \texttt{MadGraph5\_aMC@NLO} at leading order 
and passed through \texttt{Pythia}~\cite{Sjostrand:2006za} to perform showering and hadronization. 
\texttt{Delphes 3.5.0 }~\cite{deFavereau:2013fsa}  with FCC-hh detector card \footnote{\url https://github.com/delphes/delphes/blob/master/cards/FCC/FCChh.tcl}  is used 
for detector simulation. 
The jet finding is done using \texttt{FastJet}~\cite{Cacciari:2011ma} 
using an anti-$k_t$ algorithm with a distance parameter of 0.4 \cite{Cacciari:2008gp} considering the particle-flow reconstruction approach as described in 
Ref.\cite{deFavereau:2013fsa}.
Several signal samples are generated with ALP masses from 1 MeV to 150 MeV and $f_a$ is taken to be $1~\text{TeV}$.
Based on the final state, events are selected by applying the following requirements: 
\begin{itemize}
\item{only one isolated charged lepton ($e,\mu$) with $p_{\rm T} \ge 30$ GeV and $|\eta| \le 2.5$. 
Events containing additional charged leptons with $p_{\rm T} \ge 10$ GeV 
that fulfill loose isolation criteria are discarded. 
Isolated leptons are
chosen with the help of the isolation variable} $I_{\rm Rel}$ according to the
definition given in Ref. \cite{deFavereau:2013fsa}. Similar to Ref.\cite{CMS:2017dcx}, $I_{\rm Rel}$ is required to be  less than $0.15$ for muons
and $0.035$ for electrons. For loose electrons (muons), $I_{\rm Rel}$ is required to be less than $0.126 (0.25)$. 
\item{at least three jets with $p_{\rm T} \ge 30$ GeV and  $|\eta|  \leq 2.5$ from which one must be tagged as a b-jet. 
B-jet identification is based on a parametric approach which relies on
 Monte-Carlo generator information. The probability for b-jet identification is according to the parameterization of the b-tagging
efficiency available in the FCC-hh detector card.  For a jet with $ 10 < p_{T} < 500$ GeV and $|\eta| < 2.5$, the b-tagging efficiency is
taken to be $82\%$ and misidentification rates are $15\%$ and $1\%$ for c-quark jets and light flavor jets, respectively. }
\item{ the magnitude of missing transverse momentum to be greater than 160 GeV.}
\end{itemize}
For further reduction of $t\bar{t}$ and $W+jets$ backgrounds,
the transverse mass
$$M_{\rm T} = \sqrt{2p_{\rm T,l}E_{\rm T}^{\rm miss}(1-\cos\Delta\phi(\vec{p}_{\rm T,l},\vec{E}_{\rm T}^{\rm miss}))},$$
has to be greater than 160 GeV.  Moreover, 
the magnitude of the vector sum of all jets with $p_{\rm T} > 20$ GeV and $|\eta| < 5.0$, $H_{T}$, 
is required to be larger than 120 GeV.
To suppress the contribution of SM $t\bar{t}$ background, a lower cut value of 200 GeV is applied on the $M_{T2}^{W}$ variable. 
$M_{T2}^{W}$ variable has been introduced in Ref.\cite{Bai:2012gs} in searches for supersymmetric partner of the 
top quark.  
To assure the validity of the considered effective
Lagrangian, it is required that its suppression scale $f_{a}$ to be
larger than the typical energy scale of the process. Therefore, in each event the energy scale of the process
$\sqrt{\hat{s}}$ has to be much less than $f_{a}$. In this work the ALP appears as missing momentum
and $\sqrt{\hat{s}}$ is not totally measurable.
As a result, to provide the validity of the effective theory, $f_{a}$ is compared to the magnitude of  missing transverse momentum.
The magnitude of missing transverse momentum is required to be less than $f_{a}$ in each event. 
The signal efficiency after the  cuts  is found to be $12.7\%$ for the case of $m_{a} = 1$ MeV. 
The total number of background events after the cuts corresponding to an integrated luminosity of 
30 ab$^{-1}$ is $2.12\times 10^{7}$.
The signal and backgrounds efficiencies after lepton and jets selection and the cuts
on $M_{\rm T}$, $M_{T2}^{W}$, and  $H_{T}$ are presented in Table \mbox{\ref{cuteff}}.

\begin{table}[]
	\renewcommand{\arraystretch}{1.3}
	\resizebox{\textwidth}{!}{%
		\begin{tabular}{|c|c|c|c|c|c|c|c|}
			\hline
Cut                                                                         &  $c_{gg}/f_a = 0.1$ TeV$^{-1}$  &    $ c_{a\Phi}/f_a = 0.1$ TeV$^{-1} $  & $t\bar{t}$ & Single top & W+jets    & Z+jets                      & Diboson \\ \hline
Lepton and jet selection, $M_T$, MET, $H_T$, $M^W_{T2}$       & $12.7\%$   & $3.7\%$ & $0.0077\%$ & $0.0044\%$ & $5.65\times10^{-6}\%$ & $5.78\times10^{-6}\%$ & $0.0046\%$ \\ \hline
		\end{tabular}  }
	\caption{Efficiency of cuts for two signal cases with ($c_{gg}/f_a = 0.1$ TeV$^{-1}$, $m_a$ = 1 MeV);  ($c_{a\Phi}/f_a = 0.1$ TeV$^{-1}$, $m_a$ = 1 MeV);
	and for background processes after lepton and jets selection and applying cuts
	on $M_T$, MET, $H_T$, $M^W_{T2}$.}
	\label{cuteff}
\end{table}

\begin{figure}[H]
\centering
\includegraphics[width=9cm]{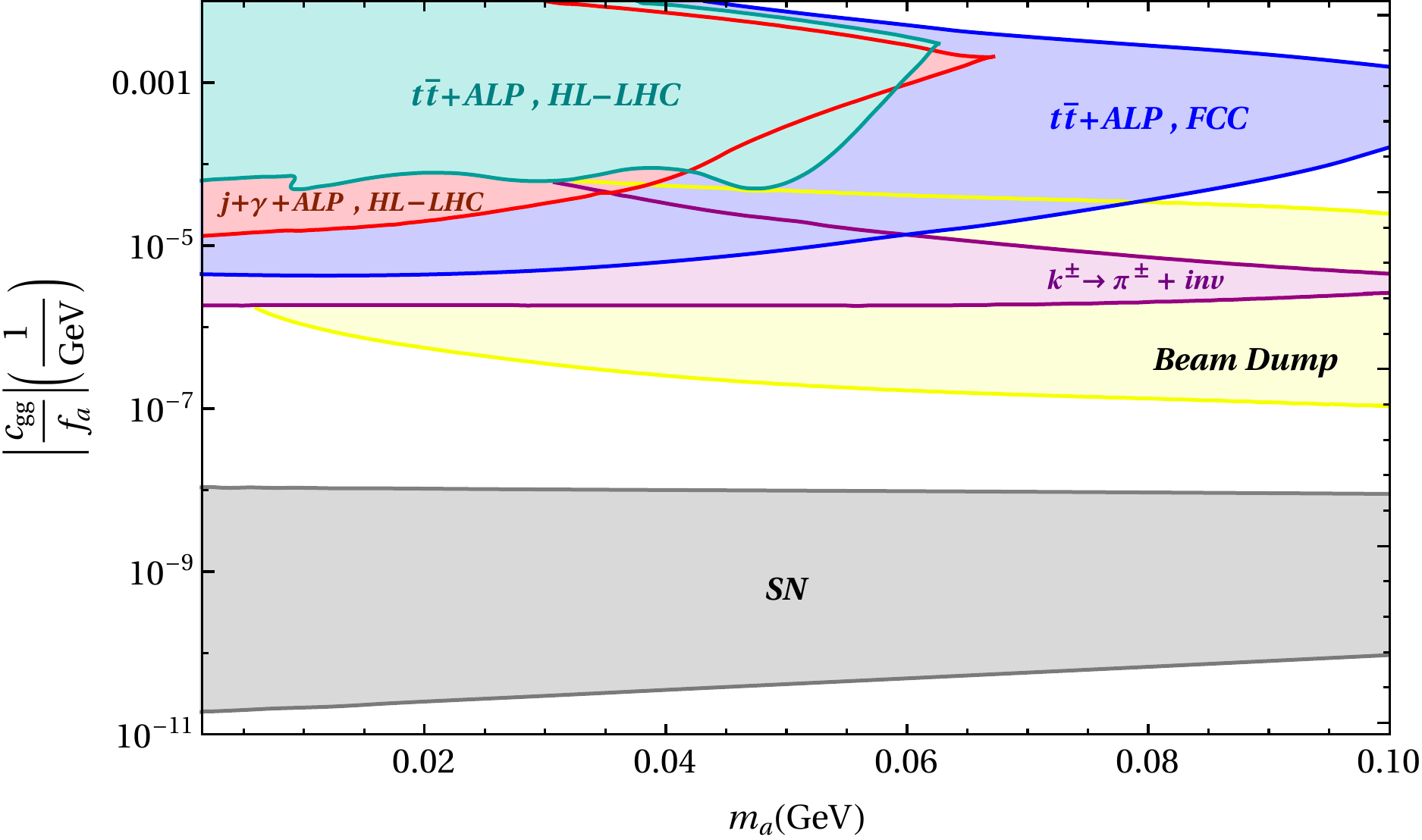}
\caption{The expected excluded regions of the
ALP model parameter space ($|c_{gg}/f_{a}|, m_{a}$) at $95\%$ CL obtained from
$t\bar{t}+ALP$ and $j+\gamma+ALP$ channels are presented. The regions derived from $t\bar{t}+ALP$ and $j+\gamma+ALP$ processes
at HL-LHC are corresponding to an integrated luminosity of 3 ab$^{-1}$ and are adapted from Ref. \cite{mojtaba}. The blue region shows the constraint obtained in the present 
analysis using $t\bar{t}+ALP$ process at FCC-hh at a center-of-mass energy of 100 TeV with 
an integrated luminosity of 30 ab$^{-1}$.
The grey region denoted by
SN presents the bound from supernova neutrino burst duration
adapted from Ref. \cite{Fukuda:2015ana}. The  region labeled by $K^{\pm} \rightarrow \pi^{\pm}+inv$ adapted from \cite{Izaguirre:2016dfi} (purple) and
beam dump (yellow) present the constraints from Kaon decay and from the
proton beam dump experiment CHARM adapted from Ref. \cite{Dobrich:2015jyk}.\label{fig2}}
\end{figure}

In order to constrain $c_{XX}/f_a$ coupling,  the first step is to set upper limit on signal cross section.
The expected upper $95\%$ CL limit on the signal cross section in the background-only hypothesis
is obtained using the standard Bayesian approach \cite{Bertram:2000br}. 
Comparing the upper bound on the signal cross section with the theoretical cross section, 
the $95\%$ CL upper limits on $|c_{XX}/f_{a}|$ are derived.
The expected $95\%$ CL upper bound on $|c_{gg}/f_{a}|$ for $m_{a} = 1$ MeV is found to be:
\begin{eqnarray}
|\frac{c_{gg}}{f_{a}}| \le  0.00446 ~\text{TeV}^{-1} ~ @ ~30 ~\text{ab}^{-1}, 
\end{eqnarray}
The prospect at HL-LHC for $m_{a} = 1$ MeV is \cite{mojtaba}:
$| \frac{c_{gg}}{f_{a}}| \le  0.063   ~\text{TeV}^{-1}  ~ @ ~3000 ~\text{fb}^{-1}$.  
Excluded regions in the  $(|c_{gg}/f_{a}|, m_{a})$ plane at $95\%$ CL 
from $t\bar{t}+ALP$ are presented in Fig.\ref{fig2}. The regions are
corresponding to integrated luminosities of 3000 fb$^{-1}$ for the LHC and  30 ab$^{-1}$
for the FCC-hh at the center-of-mass energies of 14 and 100 TeV, respectively. 
For the case of non-vanishing $c_{a \Phi}$ coupling, using the related signal and backgrounds efficiency of Table \mbox{\ref{cuteff}}, 
the upper bound on $|c_{a \Phi}/f_{a}|$  for $m_{a}$ = 1 MeV is found to be $0.11$ TeV$^{-1}$. This limit is one order of magnitude looser than
the one derived on $|c_{gg}/f_{a}|$ which is due to weaker dependence of signal cross section on $c_{a \Phi}/f_{a}$ than $c_{gg}/f_{a}$. 
The analysis does not have sensitivity to $|c_{gg}/f_a|$ greater than about 10$^{-3}$ since in this region the ALP
 will decay inside the detector and this is in contrast to our assumption of ALP being long-lived and detected as missing energy. 
 Also, for heavier ALP, its decay length tends to zero and consequently, it will decay inside the detector.
It is notable that the limits are obtained considering only statistical uncertainty.
In case of including systematic uncertainties similar to Ref.\cite{CMS:2017dcx}, the upper limit on $|c_{gg}/f_{a}|$ is weakened by about
 $2.2\%$.

\section{Discussion}
\label{sum}

ALPs are CP odd scalar particles arising from spontaneously broken global $U(1)$
symmetries which can address some of the SM shortcomings, such as strong CP problem,
baryon asymmetry, neutrino mass, and dark matter.  The potential of  $t\bar{t}+ALP$ production
to  probe  parameter space of light ALPs at FCC-hh is studied. In general, light ALPs have long lifetime
and do not decay inside the detector appearing as missing momentum in the final state. 
For the ALP mass $m_{a} = 1$ MeV, the obtained upper limit on ALP coupling with gluons $|c_{gg}/f_{a}|$ at 
FCC-hh is found to be $0.00446 ~\text{TeV}^{-1}$. This
bound corresponds to the ultimate integrated luminosity the FCC-hh will eventually operate 
based on the benchmark specifications. 
As seen in Figure \ref{fig2}, the limit on  $|c_{gg}/f_{a}|$
varies slightly as the ALP mass increases. 
In order to compare the limits obtained in this analysis with
those already derived at HL-LHC, the expected upper limits on
$|c_{gg}/f_{a}|$ at $95\%$ CL from $t\bar{t}+\text{ALP}$ and $j+\gamma+ALP$ are
presented in Figure \ref{fig2}. 
A comparison shows that the constraints obtained from FCC-hh
are  stronger than the limits derived from $t\bar{t}+\text{ALP}$ and $j+\gamma+ALP$ analyses 
at HL-LHC by one to three order of magnitudes depending on the ALP mass. 
Results of Figure \ref{fig2} indicate that the analysis of  $t\bar{t}+\text{ALP}$ FCC-hh is
capable to span a large area in the ALP parameter space which is not 
accessible by $K^{\pm} \rightarrow \pi^{\pm}+inv$, SN, and beam dump experiments.
It can be concluded that the $t\bar{t}+\text{ALP}$ production at FCC-hh provides an excellent 
way in exploring the light ALP physics as a significant portion of the parameter
space is accessible through this channel.

\vspace{6pt}

\section*{Acknowledgement}
The authors are
grateful to Yotam Soreq for providing data of Ref. \cite{yotam}. Authors would also like to
thank Pedro Ferreira da Silva and Efe Yazgan for inviting us to contribute to the Open Access Special Issue
Top Quark at the New Physics Frontier. 



\end{document}